\documentclass[aps,pre,amsmath,amssymb,reprint,groupedaddress,superscriptaddress]{revtex4-1}
\usepackage{graphicx}
\usepackage{dcolumn}
\usepackage[colorlinks,linkcolor=blue,breaklinks]{hyperref}
\usepackage{chngcntr}
\usepackage[mathlines]{lineno}

\usepackage[utf8]{inputenc}
\usepackage[T1]{fontenc}
\usepackage{newtxtext,newtxmath}

\begin{document}
\title{The Thermodynamic Origins of Chiral Twist in Monolayer Assemblies of Hard Rod-like Colloids}

\author{Yawei Liu}
\email[Corresponding author: ]{yawei.liu@sydney.edu.au}
\affiliation{ARC Centre of Excellence in Exciton Science, School of Chemistry, University of Sydney, Sydney, New South Wales 2006, Australia}
\affiliation{Beijing Key Laboratory of Ionic Liquids Clean Process, CAS Key Laboratory of Green Process and Engineering, State Key Laboratory of Multiphase Complex Systems, Institute of Process Engineering, Chinese Academy of Sciences, Beijing, 100190, China}

\author{Jared A. Wood}
\affiliation{ARC Centre of Excellence in Exciton Science, School of Chemistry, University of Sydney, Sydney, New South Wales 2006, Australia}
\affiliation{The University of Sydney Nano Institute, University of Sydney, Sydney, New South Wales 2006, Australia}

\author{Achille Giacometti}
\affiliation{Dipartimento di Scienze Molecolari e Nanosistemi, 
Universit\`{a} Ca' Foscari di Venezia
Campus Scientifico, Edificio Alfa,
via Torino 155,30170 Venezia Mestre, Italy}
\affiliation{European Centre for Living Technology (ECLT)
Ca' Bottacin, 3911 Dorsoduro Calle Crosera, 
30123 Venice, Italy}

\author{Asaph Widmer-Cooper}
\email[Corresponding author: ]{asaph.widmer-cooper@sydney.edu.au}
\affiliation{ARC Centre of Excellence in Exciton Science, School of Chemistry, University of Sydney, Sydney, New South Wales 2006, Australia}
\affiliation{The University of Sydney Nano Institute, University of Sydney, Sydney, New South Wales 2006, Australia}

\date{\today}

\begin{abstract}
The propagation of chirality across scales is a common but poorly understood phenomenon in soft matter. In this work, we use computer simulations to study chiral monolayer assemblies formed by hard rod-like colloidal particles in the presence of non-adsorbing polymer and characterize the thermodynamic driving forces responsible for the twisting. Simulations show that straight (achiral) rods assemble into monolayers with a spontaneous twist that is either left- or right-handed, while helical (chiral) rods lead to assemblies with preferential chiral features that depend on their handedness and curliness. The onset of chirality in these monolayers can be traced back to small clusters formed at the initial stage of the self-assembly. In these microscopic monolayers, entropy drives twisting in ways that differ from the assumptions on which existing continuum theory is built. Depending on the geometry of the constituent rods, the preferred chiral twist can be driven by entropy gain of the polymers, or of the rods, or both. In addition, the variation of the polymer entropy with twist depends on changes in both the surface area and the volume of the monolayer. Rod fluctuations perpendicular to the monolayer also play an important role in stabilising the twisting.
\end{abstract}
 
\maketitle

\section{Introduction}
Colloidal suspensions composed of anisotropic particles can undergo self-assembly that involves the propagation of chirality from the single-particle level to the macroscopic level, and so have emerged as a versatile platform for understanding this common phenomenon in soft matter~\cite{Barry2006,Tombolato2006,Greco2015,Dussi2016,Sharma2020}. Cholesteric liquid crystals are a well-known example, but there are many others. This includes the behaviour of colloidal suspensions of DNA, viruses, peptides, polysaccharides and various synthetic nanoparticles ~\cite{Siavashpouri2017,Miller2019,Nystrom2018a,Nystrom2018b,Aggeli2001,Lv2022,Dullens2020}. While discussion continues about the physical levers that can be used to control the phenomenon~\cite{Harris1999,Sharma2009,Morrow2017,Wang2017,Yeom2015,Sun2018,Chiappini2019}, its potential application in areas including optics, catalysis, and sensing is already being explored~\cite{Hentschel2017,Li2019,Hao2020}, and will likely accelerate in light of advances in the synthesis of anisotropic and chiral nanoparticles~\cite{Glotzer2007, Lee2018, Gonzalez-Rubio2020,Dullens2020}. It is therefore important to have a better understanding of the forces which control the propagation of chirality in these systems, and which can even drive changes in surface topology~\cite{Khanra2022}.

One typical such colloidal suspension is a mixture consisting of rod-like particles and non-adsorbing polymers in a good solvent. In these rod-polymer mixtures, the polymers, which behave as random coils with a radius of gyration $r_g$, can provide an effective attraction between the rods via \textit{depletion forces}~\cite{Asakura1954}, and thus drive the rods to assemble into diverse ordered structures~\cite{Lekkerkerker1994,Siavashpouri2019}. For example,  two-dimensional colloidal membranes can form in a suspension of filamentous viruses and dextran~\cite{Barry2010}. These colloidal membranes are liquid-like monolayer assemblies, and often have a round-shaped edge in which the constituent viruses are twisted and exhibit a chiral distribution of their orientations~\cite{Gibaud2012,Gibaud2017}. 

This chiral twist is characteristic of these nearly two-dimensional systems and is very different from the more common cholesteric twist observed in bulk (i.e., three dimensional) chiral assembly~\cite{Straley1976}. The former one is commonly known as “double twist” to distinguish it from the cholesteric (single) twist. The double twist cannot be spatially uniform in the bulk and always occurs with other deformations, with typical examples being the twist-bend and splay-twist textures~\cite{Selinger2018}. While the driving mechanism for the cholesteric twist is relatively well understood, it remains elusive for the double twist in colloidal membranes and has so far defied a complete explanation, notwithstanding recent attempts. For instance, an entropically-motivated continuum theory has been developed to explain the experimental behaviour of these colloidal membranes~\cite{Kang2016}. Briefly, the entropy, manifested through the viruses as Frank elastic energy for the twist distortion and through the polymers as an effective surface tension for the excluded volume, drives the chiral twist of the membranes. This description further assumes that the membranes are incompressible in the continuum limit.

To test the generality of this theory, and to serve as an important complementary tool to interpret experimental results, it would be useful to study such membranes using a particle-based simulation approach. This would be especially useful for analyzing small clusters formed at the onset of the self-assembly process where continuum descriptions often break down. To our knowledge, however, existing simulation studies of twisted membranes have been limited to the case of achiral rods which lack intrinsic chirality~\cite{Gibaud2012}. In this work, we therefore study membranes formed by both achiral rods and chiral rod-like helices~\cite{Frezza2013} using Langevin dynamics (LD) simulations and characterise the thermodynamic driving forces responsible for the propagation of chirality in these systems.

\section{Model and Method}

\subsection{Models for rod-polymer suspensions} 
The rod-polymer suspensions were described using a continuous potential model that approximates the well-known Asakura-Oosawa-Vrij (AO) model~\cite{Asakura1954,Liu2019}: (achiral) straight rods, described as \textit{hard spherocylinders}, were represented by a rigid linear chain of length $L$ consisting of overlapping hard spheres of diameter $D$ (Fig.~\ref{fig:fig01}a); chiral rods, described as \textit{hard helices}, were modeled as a set of hard spheres having diameter $D$ evenly arranged along a helical line of contour length $L$, pitch $p$ and radius $r$ (Fig.~\ref{fig:fig01}b); and the non-adsorbing polymers were modelled as spheres with diameter $d=2r_g$ that are freely interpenetrable to each other but experience a hard repulsion from the rod spheres. For simplicity, we set the diameter of polymer spheres $d=D$. In our simulations, the hard-core potential between rod-rod (rr) and between rod-polymer (rp) sphere pairs was replaced by a continuous pseudo-hard-core potential, i.e., $U^{\alpha\beta}(r)=50(50/49)^{49} \epsilon [(\sigma/r)^{50} -(\sigma/r)^{49}]$ ($\alpha\beta \in \{\text{rr, rp}\}$) truncated and shifted at $r^{\alpha\beta}_{cut}=(50/49)\sigma$, where $r$ is the centre-to-centre distance between the spheres, $\epsilon$ is the energy parameter, and $\sigma$ is the distance parameter with $\sigma=D$. Besides, for all rods used in this work, the distance between consecutive spheres is $0.5D$, which is sufficient to remove side effects associated with surface roughness (see Appendix.~\ref{appendixA}).

While an implicit polymer model for (achiral) straight rods such as that in Refs.~\cite{Savenko2006,Cherstvy2008,Patti2009} can allow us to simulate large systems, the corresponding model for helical rods is lacking and developing an accurate implicit polymer model for hard helices, especially in the case of large polymers, could be quite challenging~\cite{Wood2021}.

\subsection{Langevin dynamics simulation details} 
All LD simulations were carried out using LAMMPS~\cite{Plimpton1995} at a dimensionless temperature $k_BT/\epsilon=1$ (where $k_B$ is the Boltzmann constant and $T$ is the temperature). In the simulations, rod and polymer spheres are subjected to three forces: the conservative force $f^C$ computed via the pairwise interactions (i.e., the pseudo-hard-core potential); the friction force $f^F=-(m/\gamma)v$ with $m$ the mass, $\gamma$ the damping factor, and $v$ the velocity of the sphere; and the random force $f^R\propto\sqrt{k_BT m/(\Delta t\gamma)}$ with $\Delta t$ the time step. All simulations were performed in a box with periodic boundary conditions. The velocity-Verlet algorithm was used to integrate the equations of motion with a time step $\Delta t=0.001\tau$ where $\tau=D\sqrt{m/(k_BT)}$, and the damping factor was set to be $\gamma=1 \tau$. In all simulations, we set the masses of one polymer and one rod $m_p=m_r=m=1$. Simulations of large monolayers were performed in an isothermal-isobaric ($NPT$) ensemble. A Berendsen barostat with a time constant of $1\tau$ was applied. Most simulations were initialised with $N_r=480$ rods in a single hexagonally packed layer surrounded by $N_p=40000$ polymer spheres in a box with initial dimensions $44\times44\times21 D^3$. Initial configurations with different chiral twists were also used to confirm that only one handedness was stable. At least $10$ independent simulations with different initial configurations were performed for each rod shape, and all simulations were run for at least $5\times10^6$ steps to collect enough configurations at the equilibrium state.

\subsection{Free energy calculations} 
Simulations used for measuring changes in the free energy ($\Delta \Omega_{total}$) as a function of the twist ($\langle \psi_i \rangle$, see its definition in next section) were performed in a semi-grand canonical ($\mu_pVT$) ensemble with $N_r=2-61$ rods. During the simulations, $1000$ GCMC insertion and deletion moves were performed every $1000$ LD steps to maintain the chemical potential of the polymers ($\mu_p$). Simulations were initialised with $N_r$ rods in a single hexagonally packed layer surrounded by $\sim 2000$ polymer spheres in a box with dimensions $15\times15\times15 D^3$. The values of $\Delta \Omega_{total}$ as a function of $\langle \psi_i \rangle$  were evaluated by means of the umbrella sampling (US) method~\cite{Torrie1977}. We imposed a harmonic spring biasing potential given by $U = 0.5k [\langle \psi_i \rangle-\langle \psi_i \rangle_0 ]^2$ on the system using the Colvars package~\cite{Fiorin2013}. Here, $k$ is the spring constant, $\langle \psi_i \rangle_0$ is the desired twist, and $\langle \psi_i \rangle$ is the actual twist in the monolayer. Under the biasing potential, the monolayer is forced to stay in a pseudo-equilibrium state with $\langle \psi_i \rangle$ fluctuating around $\langle \psi_i \rangle_0$. Different twisted states can be described by a series of values with $\langle \psi_i \rangle_0 \in (\langle \psi_i \rangle_{min},\langle \psi_i \rangle_{max})$. In our simulations, $k=1 k_BT/deg^2$, $\langle \psi_i \rangle_{min}=-24^\circ$, $\langle \psi_i \rangle_{max}=24^\circ$, and the increment of $\langle \psi_i \rangle_0$ was $1^\circ$ or $2^\circ$. For each given $\langle \psi_i \rangle_0$, the system was equilibrated for $2\times10^6$ steps followed by another $2\times10^6$ steps production run in which data was accumulated every $1000$ LD steps. Finally, the WHAM algorithm~\cite{wham} was used to calculate the free energy change $\Delta \Omega_{total}$ as a function of $\langle \psi_i \rangle$ . For each monolayer, $10$ independent simulations were carried out to obtain good statistics. Meanwhile, to prevent the disassembly of small monolayers, additional spring forces were imposed on the rod to move it back when the distance from the centre of the rod to the centre of the monolayer is larger than a critical value $r_c$, where $r_c=0.75 D$ for $N_r=2$, $r_c=1.5 D$ for $N_r=7$, $r_c=3.0 D$ for $N_r=19$, $r_c=4.5 D$ for $N_r=37$ and $r_c=6.0 D$ for $N_r=61$. These critical values are larger than the equilibrium radii of the respective stable monolayers. 

\subsection{Excluded volume calculations} 
During the production stage of US simulations, we sampled configurations every $5000$ steps and computed the excluded volume (i.e., $\Delta V_{exc}$) for polymer spheres due to rods and the corresponding contributions from the volume and the surface area of the monolayer (i.e., $\Delta V_{exc}^{bulk}$ and $\Delta V_{exc}^{surf}$). For a given configuration, all rod spheres that had at least one polymer neighbour within $1.5D$ from their centre were classified as surface rod-spheres, and the rest were classified as bulk rod-spheres. To compute the excluded volumes, the whole system was divided into many small cubic bins with an edge length of $l=0.5D$. We confirmed that using a smaller value of $l$ (e.g., $l=0.25D$, see Appendix.~\ref{appendixB}) gave similar results. A bin was occupied by rods if there was at least one rod-sphere whose centre was less than $1.0D$ (corresponding to the polymer diameter) from the bin's centre, and the volume of the bin contributed to $\Delta V_{exc}^{surf}$ if all rod-spheres occupying this bin were surface rod-spheres, otherwise it contributed to $\Delta V_{exc}^{bulk}$. The final value of the excluded volume at a given $\langle \psi_i \rangle$ was averaged over all configurations collected at the corresponding $\langle \psi_i \rangle_0$. For these calculations, the Freud Python package~\cite{Ramasubramani2020} was used to analyse the simulation data.

\subsection{Suppressing perpendicular fluctuations} 
In the simulations for monolayers without rod fluctuations perpendicular to the monolayer, the centres of mass of all rods were constrained on a common plane via harmonic spring forces using a spring constant of $1000 k_BT/D^2$.

\section{Results and discussion}

\begin{figure}[tb]
	\begin{center}
		\includegraphics[width=0.45\textwidth]{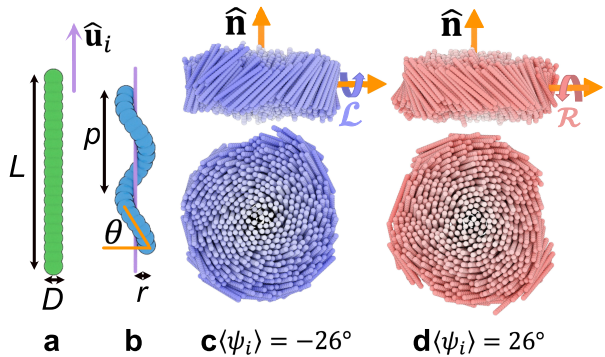}
		\caption{Spontaneous twist in monolayers of rods. (a) An achiral straight rod of length $L$ and diameter $D$. (b) A chiral rod (left-handed helix) with contour length $L$, diameter $D$, pitch $p$ and radius $r$. $\theta$ quantifies the inclination angle of the rod~\cite{Frezza2014}. (c-d) Side-view and top-view of (c) left-handed ($\mathcal{L}$) and (d) right-handed ($\mathcal{R}$) monolayers composed of straight rods with the color indicating the normalised tilted angle ($\psi_i/|\psi_i^{max}|$) between the rod axis $\hat{\textbf{u}}_i$ and the  nematic director $\hat{\textbf{n}}$.}
		\label{fig:fig01}
	\end{center}
\end{figure}

\subsection{Spontaneous twist in monolayers}
We first considered monolayers composed of (achiral) straight rods with length of $L=10 D$ (Fig.~\ref{fig:fig01}a). Figure~\ref{fig:fig01}c, d show equilibrium configurations obtained from simulations with $N_r=480$ rods surrounded by $N_p=40000$ polymer spheres at the pressure $P=1.2 k_BT/D^3$. The rods are parallel to the normal axis (i.e., the nematic director $\hat{\textbf{n}}$) at the centre, but tilt with increasing magnitude around the radial axis away from the centre. In multiple independent simulations started with an untwisted configuration, the monolayer shows nearly equal probability to end up showing left-handed ($\mathcal{L}$) or right-handed ($\mathcal{R}$) twist. Such monolayers, with roughly square edge profiles, are also predicted by the continuum theory and have been observed in experiments for small colloidal membranes~\cite{Kang2016}. 

To evaluate the degree of twist in the monolayers, we used the average tilt angle of rods with respect to the nematic director, defined as
\begin{equation}\label{eq:eq01}
	\langle \psi_i \rangle = \left\langle \dfrac{\hat{\textbf{r}}_{i} \cdot (\hat{\textbf{n}}\times \hat{\textbf{u}}_i)}{\vert\hat{\textbf{r}}_{i} \cdot (\hat{\textbf{n}}\times \hat{\textbf{u}}_i)\vert} \cos^{-1}(\vert \hat{\textbf{n}} \cdot \hat{\textbf{u}}_i \vert) \right \rangle,
\end{equation}
where $\hat{\textbf{r}}_{i}$ is unit vector connecting the center-of-mass of the monolayer and the center-of-mass of rod $i$, $\hat{\textbf{u}}_i$ is the a unit vector along the long axis of the rod and $\langle \dots \rangle$ indicates an average over all rods in the monolayer  and all configurations collected at the equilibrium state. $\langle \psi_i \rangle$ is negative for the $\mathcal{L}$ twist and positive for the $\mathcal{R}$ twist (Fig.~\ref{fig:fig01}c, d).

\subsection{Phase diagram of chirality in monolayers}
\begin{figure}[tb]
	\begin{center}
		\includegraphics[width=0.45\textwidth]{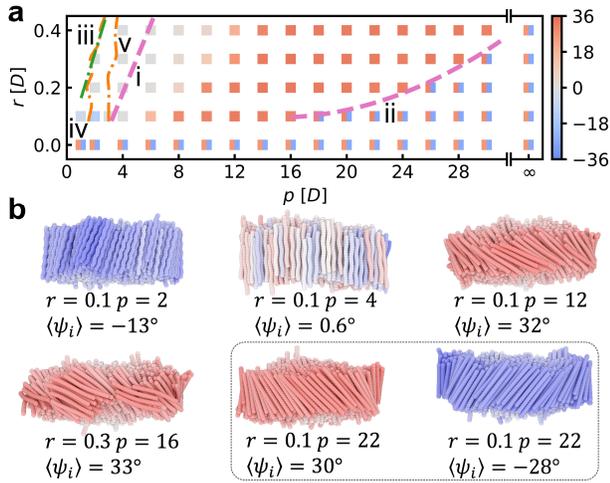}
		\caption{Phase diagram of chirality in monolayers. (a) The phase diagram shows the chirality of monolayer assemblies of hard rods with $L = 10$ and varying $r$ and $p$ at $P=1.2$. All symbols are colored according to $\langle \psi_i \rangle$ (Eq.~\ref{eq:eq01}) of the corresponding monolayer. The handedness of the monolayer with respect to that of the constituent hard rods can be the same (blue square), opposite (red square), or mixed with either $\mathcal{R}$ or $\mathcal{L}$ (bi-colored square). Note that all symbols with $r=0$ or $p=\infty$ represent the case of straight rods. Lines (i) and (ii) indicate the approximate phase boundaries for monolayers. Lines (iii-v), provided for comparison, indicate the phase boundaries between the same and opposite regions for bulk cholesteric phases. Line (iii) is given by the critical inclination angle $\theta=45^\circ$~\cite{Tombolato2006,Cherstvy2008}, while Lines (iv-v) were obtained using density functional theory at (iv) low and (v) high volume fractions, respectively~\cite{Belli2014}. (b) Typical snapshots of stable monolayers obtained from simulations using different left-handed helices.}
		\label{fig:fig02}
	\end{center}
\end{figure}

We then studied monolayers of left-handed hard helices with $L=10$ and varying $r$ and $p$ (Fig.~\ref{fig:fig01}b). A summary of the results obtained from simulations with $N_r=480$ at $P=1.2k_BT/D^3$ is reported in Fig.~\ref{fig:fig02}. In the phase diagram (Fig.~\ref{fig:fig02}a), we can identify the values of $r$ and $p$ that give rise to a chiral twist, whose handedness with respect to that of the constituent rods is (i) the same (e.g., $r=0.1$, $p=2$), (ii) the opposite (e.g., $r=0.1$, $p=12$), or (iii) mixed with either $\mathcal{R}$ or $\mathcal{L}$ (e.g., $r=0.1$, $p=22$) (Fig.~\ref{fig:fig02}b).

In Fig.~\ref{fig:fig02}a, Line (v) is the phase boundary between same and opposite regimes for the corresponding cholesteric phases at high volume fractions obtained using density functional theory~\cite{Belli2014}.  We can see that, in comparison, the corresponding phase boundary for monolayers [i.e., Line (i)] is shifted toward larger values of $p$ at $r>0.1$. Such shifting is also observed in cholesteric phases when the packing density of helices increases [compare Lines (iii)/(iv) to Line (v) in Fig.~\ref{fig:fig02}a]. Thus the difference between Line (i) and Line (v) is likely due to the higher rod packing fraction in the monolayers (0.6-0.7) compared to in the bulk cholesteric phases (0.35-0.5)~\cite{Belli2014}.

For most helices, however, their monolayer assemblies and cholesteric phases have the same handedness, supporting the experimental observation of consistent chirality between the two for rod-shaped viruses~\cite{Gibaud2017}. For weakly curled helices in the mixed regime, the monolayers can be either $\mathcal{R}$ or $\mathcal{L}$, while the cholesteric phases may only exhibit weak opposite handedness~\cite{Belli2014}. As will be elaborated later, the driving mechanism of this chiral monolayer assembly is very different from the bulk cholesteric chiral assembly that was originally predicted by Straley~\cite{Straley1976} and recently confirmed by density functional theory~\cite{Frezza2014,Belli2014} and numerical simulations~\cite{Cinacchi2017}. 

From the phase diagram, we also can see that the degree of twist (i.e., $\langle \psi_i \rangle$) is a non-monotonic function of the intrinsic pitch of the rods (i.e., $p$), which is consistent with the behaviour of bulk cholesteric phases formed by hard helices~\cite{Frezza2014,Cinacchi2017}. Starting at $p=\infty$ (i.e., straight rods), the magnitude of $\langle \psi_i \rangle$ increases as $p$ decreases, and reaches a maximum for moderately curled helices (e.g., $r=0.1$, $p=12$ and $r=0.3$, $p=16$) in the opposite regime, before decreasing to $0$ for helices at the phase boundary between same and opposite regimes (e.g., $r=0.1$, $p=4$) (Fig.~\ref{fig:fig02}b). In the same regime, $|\langle \psi_i \rangle|$ is small, but our results at $r=0.1$ show that here again $|\langle \psi_i \rangle|$ first increases and then deceases as $p$ decreases (Fig.~\ref{fig:fig02}a).

\subsection{Thermodynamic origins of chiral twist}
To study the thermodynamic origins of chiral twist in these monolayer assemblies, we considered a monolayer of $N_r$ rods in a sea of polymer spheres at fixed volume $V$ and temperature $T$. This system was kept in osmotic equilibrium with a large reservoir containing the pure polymer solution at fixed fugacity $z_p= \exp(\mu_p/k_BT)$ where $\mu_p$ is the polymer chemical potential. The grand potential of the system can be written as $\Omega_\text{total}(N_r,V,T,\mu_p) = F_r-z_p (V-V_{exc}) k_BT$, where $F_r$ is the Helmholtz energy of the rods and $V_{exc}$ is the volume excluded to the polymers by the hard rods ~\cite{Lekkerkerker1994,Bolhuis1997}. The second term on the right is the free energy of the polymers $\Omega_p$. $V_{exc}$ can be further divided into a bulk term and a surface term associated to the volume and surface area of the monolayer, respectively. Thus, we obtain the change in free energy expressed as
\begin{equation}\label{eq:eq02}
	\begin{aligned}
	\Delta \Omega_{total} & = \Delta F_r + \Delta \Omega_{p} \\
	& = \Delta F_r + z_p k_BT  \Delta V_{exc} \\
	& = \Delta F_r+z_p k_BT (\Delta V_{exc}^{bulk} + \Delta V_{exc}^{surf}).
	\end{aligned}
\end{equation}
Both $\Delta F_r$ and $\Delta V_{exc}$ depend on the twisting state of rods in the monolayer. We measured $\Delta \Omega_{total}$ as a function of $\langle \psi_i \rangle$ in a semi-grand canonical ($\mu_pVT$) ensemble with fixed $N_r$ at $z_p=1.2$ (corresponding to $P=1.2 k_BT/D^3$ in the previous simulations). The twist was constrained using the US approach, while $\Delta V_{exc}^{bulk}$ and $\Delta V_{exc}^{surf}$ were numerically calculated.

We performed a series of US simulations to calculate the changes in free energy as a function of the twist for monolayers formed by $N_r=2-61$ rods with varying $p$ (see Appendix~\ref{appendixC}). The average twist of stable monolayers monotonically increases as the monolayer size increases, which is qualitatively consistent with the theoretical description for small colloidal membranes in which the tile angles of rods at the edge have yet to reach the limiting value of $90^\circ$~\cite{Kang2016}. More importantly, we found that simulating tens of rods is sufficient to capture the chiral behaviour exhibited by the large monolayers shown in Fig.~\ref{fig:fig02}, indicating that the chirality of these monolayers is determined already during the onset of the self-assembly process.

\begin{figure}[tb]
	\begin{center}
		\includegraphics[width=0.45\textwidth]{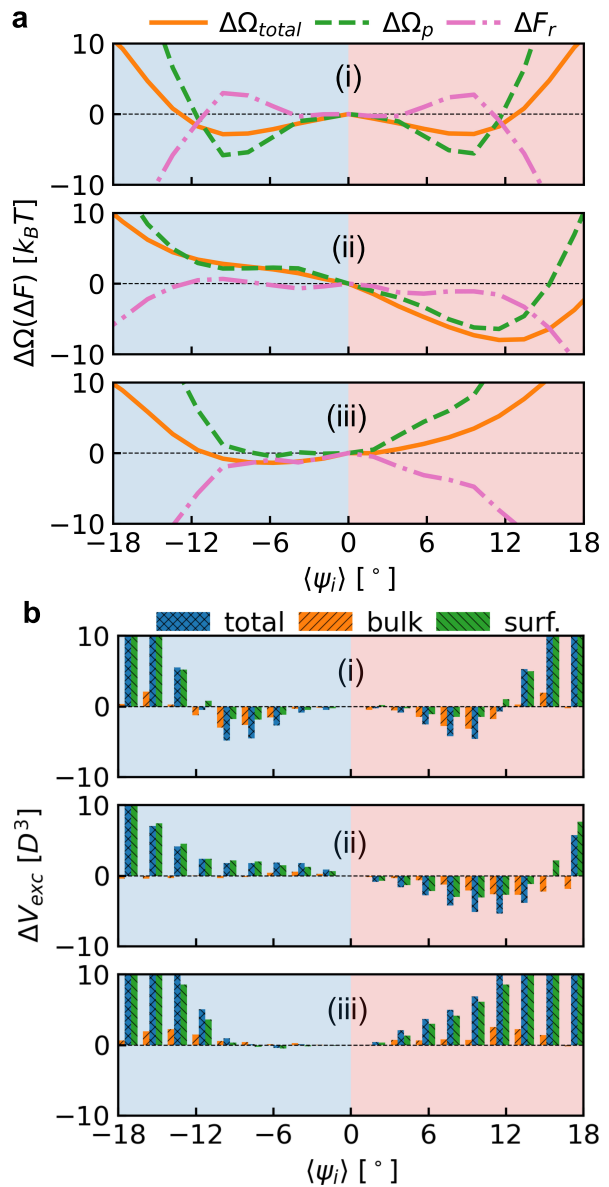}
		\caption{The thermodynamic origins of chiral twist in monolayers of different rods. (a) The changes in free energy ($\Delta \Omega_{total}$, $\Delta F_r$ and $\Delta \Omega_{p}$) as a function of the twist ($\langle \psi_i \rangle$) for monolayers formed by $N_r=37$ (i) straight rods ($r=0.0$, $p=\infty$), (ii) left-handed moderately curled helices ($r=0.1$, $p=12$), and (iii) left-handed highly curled helices ($r=0.1$, $p=2$). (b) The corresponding changes of excluded volume ($\Delta V_{exc}$, $\Delta V_{exc}^{bulk}$ and $\Delta V_{exc}^{surf}$).}
		\label{fig:fig03}
	\end{center}
\end{figure}

Figure~\ref{fig:fig03}a shows $\Delta \Omega_{total}$ \textit{vs.} $\langle \psi_i \rangle$ for three typical monolayers made up of $37$ straight rods or left-handed helices. For the monolayer of straight rods (i.e., $r=0$), the two identical minima at $\langle \psi_i \rangle>0$ and $\langle \psi_i \rangle<0$ in the curve of $\Delta \Omega_{total}$ indicate the stable twist is equally likely to be $\mathcal{R}$ or $\mathcal{L}$. For the monolayer of helices, only one local minimum appears in the curve of $\Delta \Omega_{total}$ and is located at $\langle \psi_i \rangle>0$ for the left-handed moderately curled helices (i.e., $r=0.1$, $p=12$) and at $\langle \psi_i \rangle<0$ for the left-handed highly curled helices (i.e., $r=0.1$, $p=2$), consistent with the behaviour of the larger monolayers summarised in Fig.~\ref{fig:fig02}a. 

The decomposition of $\Delta \Omega_{total}$ in Eq.~\ref{eq:eq02} reveals that chiral twist in these monolayers is stabilised by different driving forces depending on the rod shape. As shown in Fig.~\ref{fig:fig03}a (i), the twist in monolayers of straight rods is driven by the entropy gain of the polymers with respect to the untwisted state (i.e., the decease in $\Delta \Omega_p$), but further twisting beyond the equilibrium state is also prevented by the rapidly increasing entropy loss of the polymers at larger $|\langle \psi_i \rangle|$. The rod entropy in this case shows an almost opposite dependence on $\langle \psi_i \rangle$, but the polymer entropy dominates and stabilises the twist in the monolayer. For the monolayer of moderately curled helices [Fig.~\ref{fig:fig03}a (ii)], the polymer entropy also dominates and leads to a single stable twist, but now entropy gain from the rods also contributes. In sharp contrast, for the monolayer of highly curled helices [Fig.~\ref{fig:fig03}a (iii)], the single stable twist is entirely driven by the rod entropy, competing against the entropy loss of the polymers.

The polymer entropy is related to changes in the volume excluded to polymers ($\Delta V_{exc}$), which is determined by both the volume and the surface area of the monolayer. Figure~\ref{fig:fig03}b shows the change in $\Delta V_{exc}$ and its volume/surface components when twisting the three monolayers discussed in the previous paragraph. This reveals that not only the surface area but also the volume of the monolayers changes significantly during the twisting process. Especially for monolayers of straight rods [Fig.~\ref{fig:fig03}b (i)], the decrease in the volume acts as the major driving force for twisting. 

\subsection{Comparison with continuum theory} 
Having shown that different entropy components can drive rod monolayers to twist, we now compare our results with the continuum theory developed to describe such colloidal membranes~\cite{Kang2016}. The continuum theory is based on a relatively simple physical picture (see Appendix~\ref{appendixD}): that the twist is driven mainly by the entropy gained by the polymers when the membrane surface area is minimised at constant membrane volume. In this model, the polymer entropy is invariant under chirality inversion and does not contribute to the preference of the handedness, regardless of the chirality of the rods. The preferred handedness is instead attributed to an entropy term in the Frank elastic energy of the rods, whose magnitude depends on the preferred twist wavenumber that implicitly contains the chiral features of the rods.

In contrast, our simulation results reveal the existence of more complex thermodynamic behaviour. First, while the polymer entropy often drives twisting, it can also oppose twisting, with the rod entropy instead driving twisting in those cases [e.g., Fig~\ref{fig:fig03}a (iii)]. Second, the polymer entropy is asymmetric under chirality inversion for monolayers of helical particles and contributes to the preference of the handedness in these cases [e.g., Fig~\ref{fig:fig03}a (ii)]. This indicates that, at best, the Frank elastic energy in the continuum theory can depend on polymer concentration. Third, the constant-volume assumption in the continuum theory clearly breaks down, at least for the small assemblies considered here, indicating that the variation of the polymer entropy involves contributions from not only the surface area but also the volume of the monolayer (Fig.~\ref{fig:fig03}b).

Recent theoretical and experimental work~\cite{Chaturvedi2020,Kamien2020,Miller2020} has also concluded that the volume change upon twist plays a crucial role in determining the geometry and stability of colloidal membranes of rod-like particles. The geometric frustration between double-twist and splay causes the twisted monolayer to have a hyperbolic edge (i.e., "splay-twist" texture~\cite{Chaturvedi2020}), and the splay of rods away from the monolayer midplane leads to a local volume expansion, which is most significant at the top and the bottom of the monolayer edge. Based on this geometric argument, using a combination of experiments and theory in which the variation of rod density is considered, Miller et al.~\cite{Miller2020} demonstrated that for the colloidal rafts in the membranes composed of rigid rods of different lengths, the splay deformation causes expansion and compression of the inner and outer raft edges,respectively, and their competition results in spontaneous twist even for achiral systems and non-monotonic dependence of the stable twist as a function of the raft size. As for our simulated monolayers which are assembled from monodisperse rods in non-adsorbing polymers, the diffuse interfacial region exhibits a clear decline of the rod density away from the midplane, especially at the edges (see Fig.~\ref{fig:figE1}a in Appendix~\ref{appendixE}), but we
did not observe an obvious hyperboloid-like shape. Only when the perpendicular fluctuations of the rods are suppressed and their centers are confined at the 2D midplane (which is the same as the theoretical model in Refs.~\cite{Miller2020}), does our monolayer exhibit a hyperboloid-like shape (see Fig.~\ref{fig:figE1}b in Appendix~\ref{appendixE}). Moreover, in our systems, the total volumes of small monolayers (which are made up of $37$ rods in our US simulations) are always decreasing during the initial twisting process (see Fig.~\ref{fig:figE2} Appendix~\ref{appendixE}), suggesting that the volume expansion due to the splay deformation does not dominate, at least in these small monolayers. All these results suggest that the volume change due to twist is important to the stable texture in various colloidal membranes of rod-like particles and thus should not be ignored in theoretical models.

 \begin{figure}[tb]
	\begin{center}
		\includegraphics[width=0.45\textwidth]{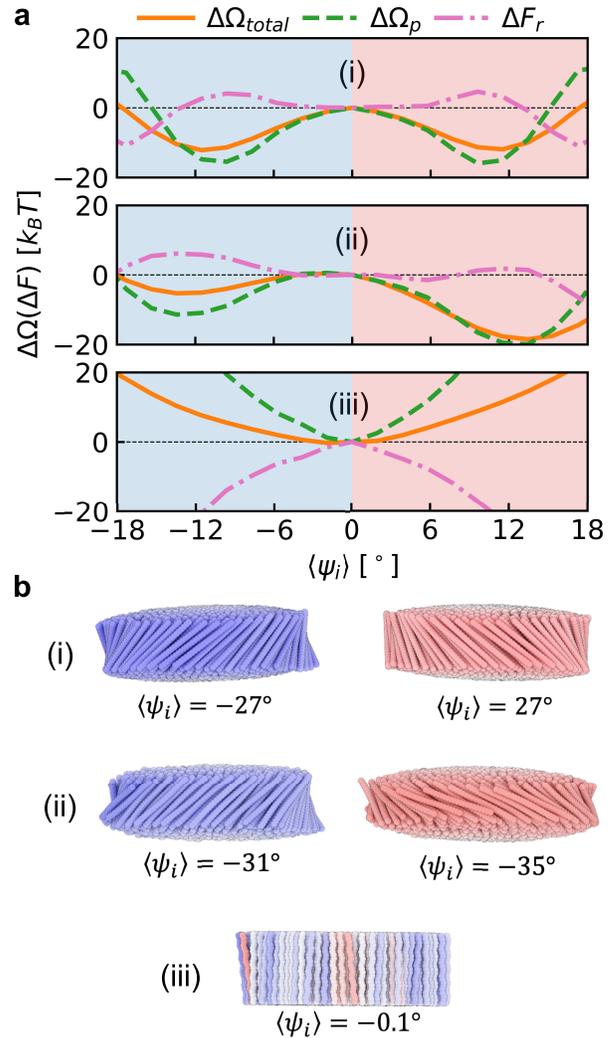}
		\caption{The role of perpendicular fluctuations. (a) The changes in free energy ($\Delta \Omega_{total}$, $\Delta F_r$ and $\Delta \Omega_{p}$) as a function of the twist ($\langle \psi_i \rangle$) for monolayers formed by $N_r=37$ (i) straight rods ($r=0.0$, $p=\infty$), (ii) left-handed moderately curled helices ($r=0.1$, $p=12$), and (iii) left-handed highly curled helices ($r=0.1$, $p=2$) when the fluctuations perpendicular to the plane of the monolayer are artificially suppressed. (b) Typical snapshots of large equilibrated monolayers formed by $N_r=480$ rods (corresponding to those in \textbf{a}), obtained from simulations without rod fluctuations perpendicular to the monolayer.}
		\label{fig:fig04}
	\end{center}
\end{figure}

\subsection{The role of perpendicular fluctuations} 
Finally, we considered a special entropy contribution from rods related to their fluctuations perpendicular to the monolayer. Figure~\ref{fig:fig04}c shows $\Delta \Omega_{total}$ \textit{vs.} $\langle \psi_i \rangle$ for the three example monolayers when the centres-of-mass of all rods are constrained to the midplane of the monolayer. The rod fluctuations out of the plane are expected to produce surface roughness and so increase the volume excluded to the polymers. We found that suppressing the fluctuations resulted in more entropy gain for the polymers upon twisting for monolayers of straight rods and moderately curled helices [see larger changes of $\Delta \Omega_p$ in Fig.~\ref{fig:fig04}a (i) and (ii) compared to that in Fig.~\ref{fig:fig03}a (i) and (ii)]. This stabilises the twisted states for these monolayers, and even adds a new metastable twisted state for the monolayer of moderately curled helices [Fig.~\ref{fig:fig04}a (ii)]. In contrast, for the monolayer of highly curled helices, the polymer entropy increases dramatically upon twisting when the fluctuations are suppressed, causing the original weakly-stable twisted state to disappear [Fig.~\ref{fig:fig04}a (iii)]. These results are consistent with unconstrained simulations of large monolayers (Fig.~\ref{fig:fig04}b), and clearly show that rod fluctuations perpendicular to the monolayer have important effects on the stability of the chiral twists that depend on the shape of the individual rods.

We note that such contributions from rod fluctuations perpendicular to the monolayer are either ignored in continuum models of colloidal membranes or only taken into account in a simplistic manner (which is not curliness-dependent)~\cite{Kang2016} (see Appendix~\ref{appendixD}). Our simulation results, however, indicate that these fluctuations can play a crucial role in the stability of the chiral twist, and thus  may need to be accurately described in order to predict the stable chiral twist.

\section{Conclusions}
In summary, we have used a simple model to characterise spontaneous chiral twist in monolayers assembled from either achiral or chiral rods in non-adsorbing polymer solutions, and thus to reveal the thermodynamic driving forces responsible for the chirality propagation from single particles to their assemblies. Note that the chiral twist discussed in this work is the double twist, which is distinct from the cholesteric (single) twist~\cite{Selinger2018}.

Depending on the geometry of the constituent rods, their monolayer assemblies exhibit a broad range of chiral behaviour, including variations in handedness and twist magnitude. Compared to the constituent rods, the (achiral) straight rods and weakly curled helices form monolayers with either $\mathcal{R}$ or $\mathcal{L}$ twist (i.e., the mixed regime), moderately curled helices form monolayers with the opposite handedness (i.e., the opposite regime), and highly curled helices form monolayers with the same handedness (i.e., the same regime). Moreover, the degree of twist in the monolayers is a non-monotonic function of the intrinsic pitch of the helices, with the most twisted monolayers forming from moderately curled helices [Fig.~\ref{fig:fig02}a].

The thermodynamic forces responsible for spontaneous chiral twist also vary dramatically between different particle shapes. In the mixed and the opposite regimes, the twist in monolayers is mainly driven by the polymer entropy [Fig.~\ref{fig:fig03}a (i)]. As the rods becomes more curled, the rod entropy also contributes to the twist, and only the twisted state with the opposite handedness remains stable [Fig.~\ref{fig:fig03}a (ii)]. For even more curled helices, only one weakly twisted state with the same handedness is stable, and is entirely driven by the rod entropy [Fig.~\ref{fig:fig03}a (iii)].

Our simulation results also indicate important contributions from the volume change upon twist and the rod fluctuations perpendicular to the monolayer that have so far been ignored in continuum theories. Our preliminary results, obtained from Monte Carlo (MC) simulations, for rods held together by explicit attraction rather than polymer depletion indicate a similar complexity (see Appendix~\ref{appendixF}). Overall, we find increasing deviations from current continuum theory as the attractive forces holding the rods together become weaker, regardless of whether they are due to direct energetic or indirect entropic effects (see Appendix~\ref{appendixF}). All these results contribute to our understanding of chirality transmission across scales when chiral objects assemble into larger aggregates.

While our simulations were based on a simplified model for rod-like colloids (i.e., hard spherocylinders and helices), they clearly show that twisted colloidal membranes can also be formed by helical rods, which could be an interesting behaviour to investigate in future experiments by using similar natural and synthetic particles~\cite{Barry2006,Zhu2014,Feng2017,Tao2019}. Our current work also offers a helpful reference for understanding the behaviour in more complex systems. It would be very useful to consider models which are closer to the chiral rods (e.g. fd-virus and DNA origami rods) used in experiments of colloidal membranes. For example, using a “straight and helically-decorated” model~\cite{Tombolato2006,Wensink2017} would allow us to compare the computational and experimental results more directly. Meanwhile, recent experiments have shown that the shape fluctuation of chiral rods also dramatically affects their assembled structures~\cite{Tortora2020}, thus it would be interesting to consider the flexibility of rod-like particles in future work.

\begin{acknowledgements}
This work was supported by the Australian Research Council under Grant CE170100026 and by the MIUR PRIN-COFIN2017 \textit{Soft Adaptive Networks} grant 2017Z55KCW. Computational resources were provided by the Sydney Informatics Hub, a Core Research Facility of the University of Sydney.
\end{acknowledgements}

\bibliography{text}

\clearpage
\onecolumngrid

\appendix

\setcounter{figure}{0} 
\counterwithin{figure}{section}
\section{Effect of the number of spheres in one rod}\label{appendixA}
\begin{figure*}[h!]
	\begin{center}
		\includegraphics[width=0.9\textwidth]{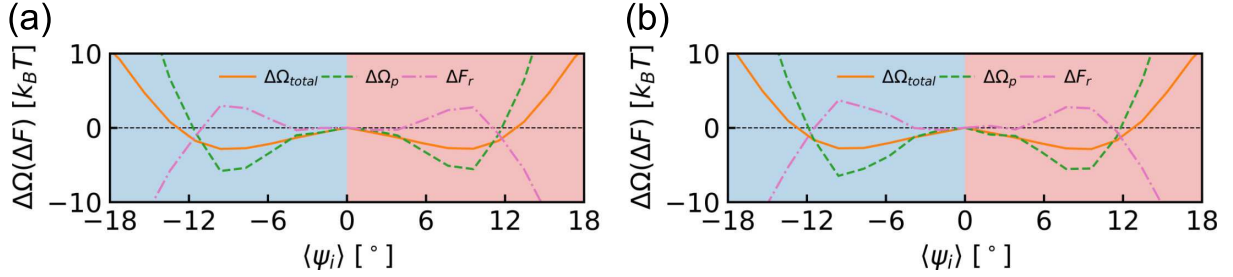}
		\caption{The total free energy ($\Delta \Omega_{total}$) and its decomposition ($\Delta F_r$ and $\Delta \Omega_{p}$)as a function of the degree of the twist ($\langle \psi_i \rangle$) for a monolayer formed by $N_r=37$ straight rods with (a) $21$ and (b) $41$ spheres in each rod.}
		\label{fig:figA1}
	\end{center}
\end{figure*}

For all results reported in the main text, the contour length $L=10$ and there are $21$ spheres evenly distributed in each rod (i.e. the distance between consecutive spheres is $0.5D$). We confirmed that this model is smooth enough to remove side effects associated with grooves between overlapping spheres. Fig.~\ref{fig:figA1} shows that using more spheres in each rod yields similar results.

\setcounter{figure}{0} 
\counterwithin{figure}{section}
\section{Effect of the bin size to determine the excluded volume}\label{appendixB}
\begin{figure*}[h!]
	\begin{center}
		\includegraphics[width=0.9\textwidth]{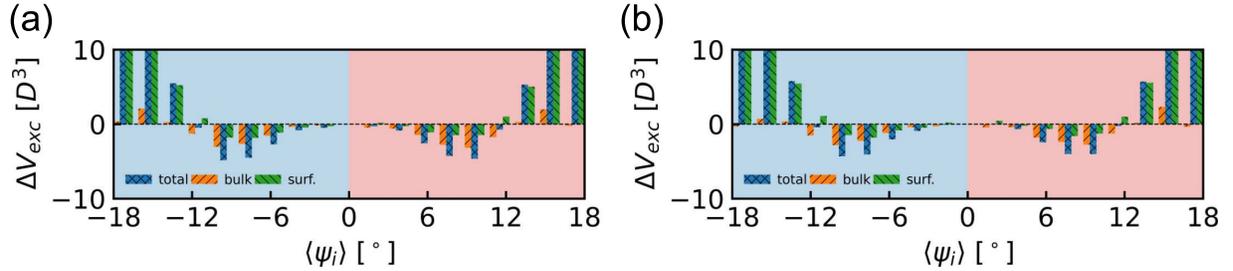}
		\caption{The change in the total excluded volume ($\Delta V_{exc}$) as a function of $\langle \psi_i \rangle$ and its decomposition into bulk and surface contributions ($\Delta V_{exc}^{bulk}$ and $\Delta V_{exc}^{surf}$) for monolayers of $37$ straight rods. In the calculations, the edge length of the cubic volumetric bins was $0.5D$ in (a) and $0.25D$ in (b).}
		\label{fig:figB1}
	\end{center}
\end{figure*}

To compute the excluded volumes during the umbrella sampling simulations, the whole system was divided into many small cubic bins with an edge length of $l=0.5D$. We confirmed that using a smaller value of $l$ (e.g., $l=0.25D$, Fig.~\ref{fig:figB1}) gave similar results.

\setcounter{figure}{0} 
\counterwithin{figure}{section}
\section{Free energy changes for monolayers composed of different numbers of rods}\label{appendixC}
\begin{figure*}[h!]
	\begin{center}
		\includegraphics[width=1\textwidth]{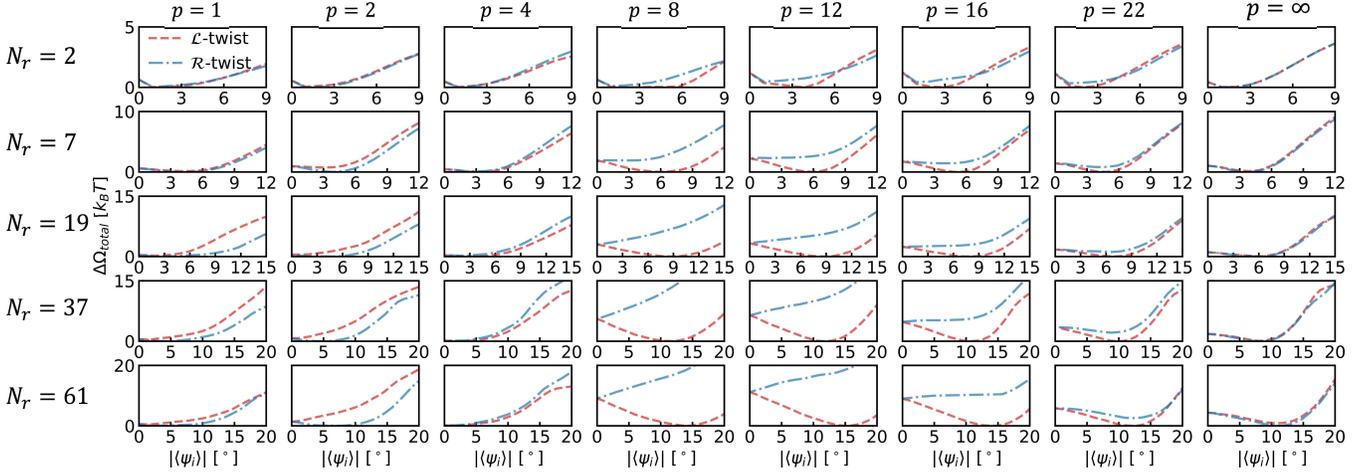}
		\caption{The total free energy change ($\Delta \Omega_{total}$) as a function of the twist ($\langle \psi_i \rangle$) for monolayers formed by $N_r=2-61$ rods with $L=10$, $r=0.1$ and varying $p$.}
		\label{fig:figC1}
	\end{center}
\end{figure*}

We performed a series of umbrella sampling simulations to calculate the change in free energy as a function of the twist for monolayers formed by $N_r=2-61$ rods with $L=10$, $r=0.1$ and varying $p$. The obtained results shown in Fig.~\ref{fig:figC1} suggest that when $N_r>=19$, the free energy minima in these curves are consistent with the phase behaviours reported in Fig.~\ref{fig:fig02} (a) in the main text, as discussed in the main text.

\begin{figure*}[h]
    \begin{center}
    \includegraphics{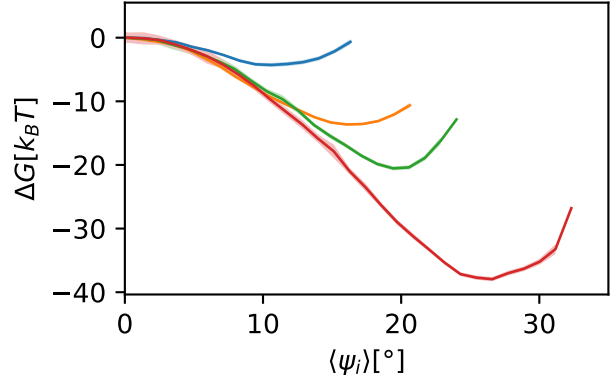}
    \caption{The change in Gibbs free energy ($\Delta G$) as a function of the twist ($\langle \psi_i \rangle$) for monolayers formed by $N_r=61$ straight rods (Blue), $N_r=120$ (Orange), $N_r=240$ (Green), and $N_r=480$ (Red). Shaded areas indicate one standard error of the mean.}
    \label{fig:figC2}
    \end{center}
\end{figure*}

We also examined the free energy of twisting for larger monolayers. For computational efficiency, this work was performed using Monte Carlo (MC) simulations of the standard hard AO interaction model\cite{Wood2021}. The same rod length and polymer size was used as in the rest of the paper, with natural twisting observed at a polymer packing fraction of $0.467$ and pressure of $1.2$. Staged umbrella sampling\cite{Meijer1994} with the same bias as in the main text was then used to determine the change in Gibbs free energy as a function of the average tilt angle for several different monolayer sizes, with the free energy profiles assembled using the MBAR technique\cite{Shirts2008}. The results are shown in Fig. \ref{fig:figC2} and are consistent with the results presented in Fig.~\ref{fig:figC1} (at bottom right) for $N_r=61$.

There are two notable changes as the size of the monolayer increases. First, the magnitude of the preferred average twist increases with the size of the cluster, despite the fraction of rods at the edge of the cluster shrinking with increasing cluster size. This indicates that the local twist at the edge has yet to reach its limiting value of 90\textdegree. Once that occurs, we expect the average twist to reach a maximum value and then to gradually decrease. Second, the depth of the free energy minimum, and consequently the free energy barrier to reversing twist direction, increases with size. This increase is roughly linear with the number of rods and monolayer area.

\setcounter{figure}{0} 
\counterwithin{figure}{section}
\section{Brief review of the continuum theory for colloidal membranes}\label{appendixD}

In the continuum theory~\cite{Kang2016b}, the colloidal membrane is treated as  a continuum medium composed of rods at constant density, and the membrane has a fixed number of rods and a constant volume. In the membrane, a rod at $\textbf{x}$ (i.e. a given position in the membrane plane) is tilted by  $\theta(\textbf{x})$  with respect to the normal director. The membrane half-thickness can be written as $h(\textbf{x})=t\cos\theta(\textbf{x})+b(\textbf{x})$ where $t$ is the half-length of the rod and $b(\textbf{x})$ is the height fluctuation amplitude of the rod. Assuming a circularly-symmetric membrane of radius $R$ and using cylindrical coordinates, one can have $h(r)$, $b(r)$, and $\theta(r)$ that only depend on the radial coordinate.

The free energy associated with the twist distortion of the rods is described by the Frank elastic free energy. Using the one-constant approximation, the free energy is given by
\begin{linenomath}
	\begin{equation}\label{eq:eqs01}
	F_\text{Frank} = 2\pi K \int_0^R dr h \left[r(\partial_r\theta)^2+\sin2\theta\partial_r\theta+\frac{\sin^2\theta}{r}-2qr\partial_r\theta-q\sin2\theta\right],
	\end{equation}
\end{linenomath}
where $K$ is the Frank elastic constant and $q$ is the preferred twist wavenumber associated with the intrinsic chirality of the constituent rods.

The free energy of the polymers is related to the volume excluded to them by the rods. For polymers small compared to the dimensions of the membrane, this excluded volume is approximately $V_0+aA$, where $V_0$ is the volume of the membrane, $A$ is the surface area of the membrane, and $a$ is the polymer characteristic radius. $V_0$ is assumed to be constant, so the corresponding free energy is similar to an effective surface tension, which is expressed as
\begin{linenomath}
	\begin{equation}\label{eq:eqs02}
	F_\text{polymer} = 4\pi n a k_B T \left[\int_0^R dr r \sqrt{1+(\partial_rh)^2}+Rh(R)\right],
	\end{equation}
\end{linenomath}
where $n$ is the polymer concentration.

If the rod fluctuations perpendicularly to the membrane plane are ignored, then $h\approx t\cos\theta$, and the profile of the membrane is can be obtained by minimizing the total free energy over $h(r)$ using a volume-conserving Lagrange multiplier $\lambda$, i.e.,
\begin{linenomath}
	\begin{equation}\label{eq:eqs03}
	F = F_\text{Frank} + F_\text{polymer} + \lambda \left[ V_0-4\pi \int_0^R dr r h \right],
	\end{equation}
\end{linenomath}

In this continuum theory, the key points for the twist are:
\begin{itemize}
	\item The preferred handedness in membranes composed of chiral rods is entirely determined by the non-zero $q$ through the free energy of the rods. For example, when $q>0$, twisted membranes with $\partial_r\theta>0$ have lower energy than those with $\partial_r\theta<0$.
	\item The polymer entropy drives the twist via minimizing the excluded volume, but it is invariant under the chirality inversion $\theta \rightarrow -\theta$, and thus does not contribute to the preferred handedness regardless of the intrinsic chirality of the rods (i.e. $q$).
	\item The excluded volume only depends on the surface area of the membrane since the volume of the membrane is assumed to be constant.
\end{itemize} 

The rod fluctuations perpendicular to the membrane plane have complicated, non-linear effects on the free energy, and so are very difficult to accurately account for in the theory. The continuum theory in Ref.~\cite{Kang2016b} only considered fluctuations of single rods and thus ignored their interactions and correlated motion. The corresponding free energy was calculated in the small rod angle and small fluctuation amplitude limit, and the rods were assumed to be packed hexagonally and to maintain a constant perpendicular distance $\xi$ between nearest-neighbors. Under these assumptions, the rod fluctuation free energy is given by
\begin{linenomath}
	\begin{equation}\label{eq:eqs04}
	F_\text{fluctuation} = \frac{8\pi^2 n a k_B T}{\sqrt{3}\xi^2}\int_0^R dr r \cos \theta \left[h-t\cos\theta-(2\pi na)^{-1/2}\right].
	\end{equation}
\end{linenomath}

\clearpage
\setcounter{figure}{0} 
\counterwithin{figure}{section}
\section{Density profiles and volume changes of assembled monolayers}\label{appendixE}

\begin{figure*}[h]
    \begin{center}
    \includegraphics{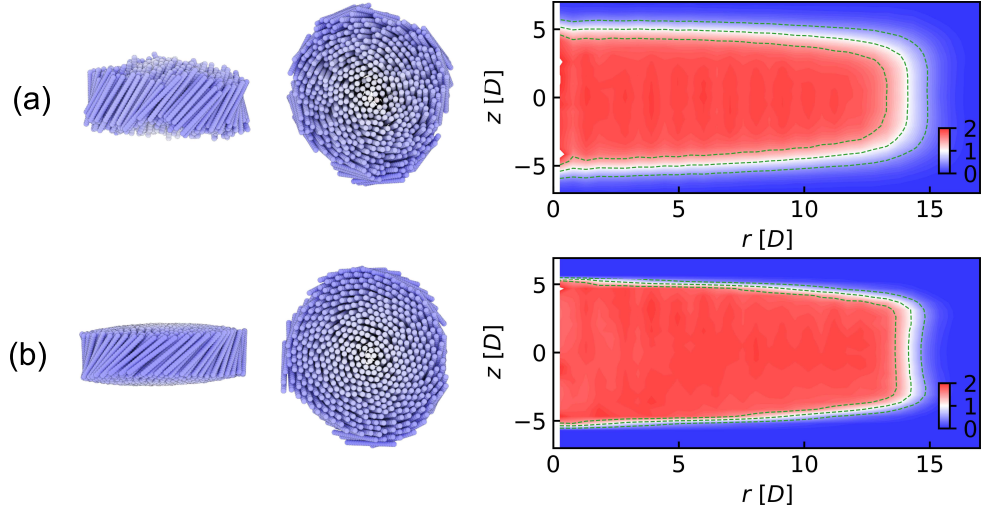}
    \caption{The snapshots and the density profile for (a) a free monolayer suspended in nonadsorbing polymers (b) a confined monolayer with centers of all rods fixed at the 2D midplane. The dashed lines are density contours at $\rho=0.5$, $1.0$, and $1.5$ $D^{-3}$. The slightly concave contour lines in (b) are the results of the hyperboloid-like shape. All monolayers are made up of (achiral) straight rods.}
    \label{fig:figE1}
    \end{center}
\end{figure*}

\begin{figure*}[h]
    \begin{center}
    \includegraphics{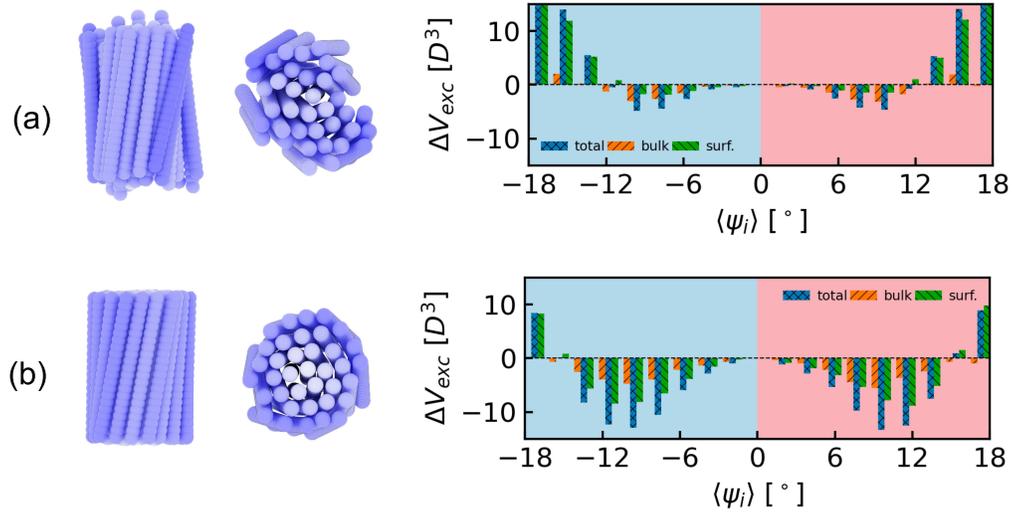}
    \caption{The changes of excluded volume for the two monolayers made up of $37$ straight rods. (a) A free monolayer suspended in nonadsorbing polymers. (b) A confined monolayer with centers of all rods fixed at the 2D midplane.}
    \label{fig:figE2}
    \end{center}
\end{figure*}

\setcounter{figure}{0} 
\counterwithin{figure}{section}
\section{Monolayers held together by explicit attraction}\label{appendixF}

We observed similar twisting behavior in the absence of polymer depletion when the rods interacted with each other via an explicit attractive potential. In this model, the rods consisted of $40$ spheres over a length $L=10$ with the spheres in different rods interacting with each other via a square well potential of width $w=1$ and depth $1/20^2=0.0025\epsilon_0$, and the temperature scaled to $T=k_{B}T/\epsilon_0$. There was no polymer in the model, with rod-attraction coming only from the square well interactions. All work was performed using monolayers of 61 rods. Using the same MC simulation and umbrella sampling approach described in section~\ref{appendixC}, this time in the $NVT$ ensemble, we calculated the Helmholtz free energy change ($\Delta F$) along with the change in the mean potential energy ($\Delta U$) as a function of the average twist. Using this, we also determined the entropy contribution to $\Delta F$ using $- T\Delta S = \Delta F - \Delta U$.

\begin{figure*}[h]
    \begin{center}
    \includegraphics{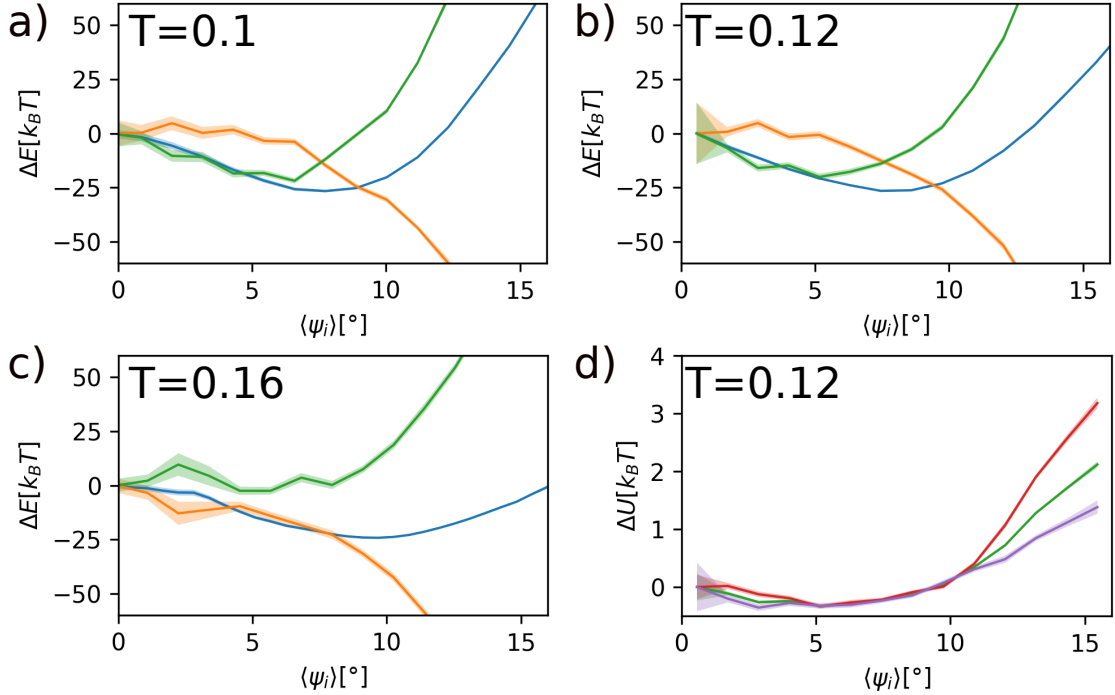}
    \caption{The change in free energy $\Delta F$ (Blue) and its decomposition into $-T\Delta S$ (Orange) and $\Delta U$ (Green) as a function of twist for monolayers formed by $N_r=61$ square-well rods at temperatures of a) $T=0.1$, b) $T=0.12$, and c) $T=0.16$. d) Shows $\Delta U$ normalized by the number of rods for all rods in the monolayer (Green), for rods located at the edge of the monolayer (Red), and for rods located in the interior (Purple). Shaded areas indicate one standard error of the mean.}
    \label{fig:figF1}
    \end{center}
\end{figure*}

The energy changes obtained as a function of the twist angle at three different temperatures are shown in Fig.~\ref{fig:figF1} (a-c). At the lower temperatures ($T=0.1$ and $T=0.12$), the initial twisting is driven entirely by the potential energy of the rods, with the rod entropy only becoming a significant driver at higher angles. This is similar to what we observed for straight rods when the twisting was driven by depletion-attraction [Fig.~\ref{fig:fig03}a (i) in the main text]. In that case, the rod free energy $\Delta F_r$, which is analogous to $-T\Delta S$ in this model, opposes twisting at small angles and only drives it at larger angles. Similarly, the force holding the monolayer together in both models ($\Delta\Omega_p$ and $\Delta U$) favors small twists but opposes larger ones.

Breaking down the potential energy into contributions from (i) rods in the interior of the monolayer and (ii) those at the edge, provides insight into whether the initial twisting is driven by minimization of the surface energy of the monolayer. This breakdown is shown for $T=0.12$ in Fig.~\ref{fig:figF1} (d). Surprisingly, we find very little difference in the potential energy change per rod between the different subgroups until past the optimum twist angle, counter-intuitively indicating that the edge of the monolayer, where the twisting is the greatest and there are fewer interactions per rod, is not more energetically advantaged or disadvantaged by the twisting than the rest of the cluster.

\begin{figure*}[h!]
	\begin{center}
		\includegraphics[width=0.9\textwidth]{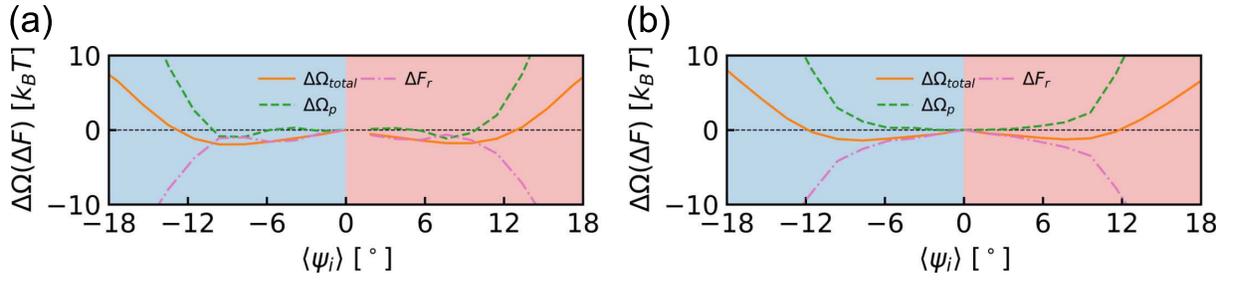}
		\caption{The change in total free energy ($\Delta \Omega_{total}$) and its decomposition ($\Delta F_r$ and $\Delta \Omega_{p}$) as a function of the twist ($\langle \psi_i \rangle$) for a monolayer formed by $N_r=37$ straight rods at a polymer fugacity of (a) $z_p=1.0$ and (b) $z_p=0.8$.}
		\label{fig:figF2}
	\end{center}
\end{figure*}

At higher temperature [$T=0.16$, Fig.~\ref{fig:figF1} (c)], the driving forces change, with the rod entropy now driving the twist at all angles and the potential energy always opposing the twist. This indicates that twisting becomes increasingly driven by the rod entropy as the monolayer density decreases. Consistent with this, we observe a similar change for depletion-driven twisting as the polymer fugacity decreases, i.e. with the rod entropy $F_r$ increasingly favoring twisting at small twists (e.g., $\mu_p=0.8$) in the depletion-driven case of straight rods [see Fig.~\ref{fig:figF2} compared to Fig.~\ref{fig:fig03}a in the main text].

\end{document}